\documentclass[letterpaper,twoside,twocolumn,english,aps]{revtex4}

\usepackage[T1]{fontenc}
\usepackage[latin9]{inputenc}
\usepackage{amsmath}
\usepackage{amssymb}
\usepackage{graphicx}

\makeatletter

\pdfpageheight\paperheight
\pdfpagewidth\paperwidth

\@ifundefined{textcolor}{}
{%
 \definecolor{BLACK}{gray}{0}
 \definecolor{WHITE}{gray}{1}
 \definecolor{RED}{rgb}{1,0,0}
 \definecolor{GREEN}{rgb}{0,1,0}
 \definecolor{BLUE}{rgb}{0,0,1}
 \definecolor{CYAN}{cmyk}{1,0,0,0}
 \definecolor{MAGENTA}{cmyk}{0,1,0,0}
 \definecolor{YELLOW}{cmyk}{0,0,1,0}
}

\makeatother

\usepackage{babel}
\begin{document}
\title{Physical Origin of the Universal Three-body Parameter in Atomic Efimov
Physics}
\author{Pascal Naidon$^{1}$\thanks{pascal@riken.jp}, Shimpei Endo$^{2}$,
and Masahito Ueda$^{2}$}
\affiliation{$\,^{1}$RIKEN Nishina Centre, RIKEN, Wak\={o} 351-0198, Japan, }
\affiliation{$\,^{2}$Department of Physics, University of Tokyo, 7-3-1 Hong\={o},
Bunky\={o}-ku, T\={o}ky\={o} 113-0033, Japan}
\date{\today}
\begin{abstract}
We address the microscopic origin of the universal three-body parameter
that fixes the spectrum of three-atom systems in the Efimov regime.
We identify it with the van der Waals two-body correlation, which
causes the three-atom system to deform when the three atoms come within
the distance of the van der Waals length, effectively preventing them
from coming closer due to the kinetic-energy cost associated with
this three-body deformation. This deformation mechanism explains the
universal ratio of the scattering length at the triatomic resonance
to the van der Waals length observed in several experiments and confirmed
by numerical calculations. 
\end{abstract}
\maketitle

\section{Introduction}

In recent years, the investigation of Efimov physics~\citep{Efimov1971,Ferlaino2010},
the universal physics of few particles interacting via nearly resonant
short-range interactions, has developed tremendously, both on the
experimental~\citep{Kraemer2006,Ottenstein2010,Huckans2009,Williams2009,Zaccanti2009,Barontini2009,Wenz2009,Pollack2009,Gross2009,Knoop2009,Lompe2010PRL,Lompe2010Science,Nakajima2010,Gross2010,Nakajima2011,Berninger2011,Ferlaino2009,Ferlaino2010,Machtey2012,Zenesini2012,Knoop2012}
and theoretical fronts~\citep{Petrov2004,Braaten2007,Yamashita2006,Hammer2007,vonStecher2009}.
The essence of this physics is the appearance of a universal $1/R^{2}$
attraction between three particles at an average separation $R$.
This long-range three-body attraction, discovered by V.~Efimov~\citep{Efimov1971},
emerges from the pairwise interactions, despite their finite range.
It can be interpreted as an interaction between two particles mediated
by a third particle. Its strength is universally determined by the
masses and quantum statistics of the particles. The Efimov attraction
extends from distances on the order of the range of the interaction
$b$ to distances on the order of $\vert a\vert$, where $a$ is the
scattering length of the pairwise interaction. It therefore requires
$\vert a\vert>b$, a condition well satisfied for resonant pairwise
interactions. At the unitarity limit $a\to\infty$, the Efimov attraction
extends to infinity. Decaying as $1/R^{2}$, it supports an infinite
number of bound states known as the Efimov trimers. Furthermore, each
bound state is related to the neighbouring state by a scale transformation,
due to the scale invariance of $1/R^{2}$ potentials~\citep{Efimov1971,Braaten2007},
so that the energy spectrum forms a geometric series. This constitutes
the most remarkable and characteristic feature of the Efimov trimers.
However, since the three particles are attracted to each other, the
physics at short separations comparable to the range $b$ fixes the
wave functions and spectrum of the Efimov trimers.

Until recently, little had been known about this short-distance physics.
Efimov's original investigation made use of the asymptotic two-body
behaviour (or the zero-range potential limit) to derive the three-body
attraction, but did not address the short-distance region directly.
Its effect on the longer-distance region was accounted for by a three-body
boundary condition, expressed either as a phase in the three-body
wave function or a log-periodic inverse length $\Lambda$ known as
the Efimov three-body parameter~\citep{Efimov1971}. This long-distance
picture is equivalent to a zero-range low-energy picture, where $\Lambda$
plays the role of the parameter that renormalises the low-energy effective
field theory~\citep{Hammer2007}. The Efimov effect is the only known
physical example of the renormalisation-group limit cycle~\citep{Wilson1971}.
Since the short-distance region involves the short-range details of
the interaction potentials, $\Lambda$ has long been thought to be
a non-universal quantity that is strongly dependent on the individual
properties of the system.

Later, it was found that $\Lambda$ is universally determined in cases
where a length scale larger than $b$ arises in the problem, most
notably in the case of a narrow Feshbach resonance in the pairwise
interaction~\citep{Chin2010}, which entails a large and negative
effective range setting the value of $\Lambda$~\citep{Petrov2004},
and the case of particles with additional dipolar interactions, whose
strength also sets the value of $\Lambda$~\citep{YujunWang2011PRL}.
In the absence of such large length scales, however, it was believed
that $\Lambda\sim1/b$, but its precise value would vary by a factor
within the entire log-period $e^{\pi/s_{0}}\approx22.7$ from one
system to another, or even from one Feshbach resonance to another
within the same system~\citep{DIncao2009}.

However, several recent experiments with identical ultra-cold atoms~\citep{Kraemer2006,Ottenstein2010,Huckans2009,Williams2009,Zaccanti2009,Barontini2009,Wenz2009,Pollack2009,Gross2009,Knoop2009,Lompe2010PRL,Lompe2010Science,Nakajima2010,Gross2010,Nakajima2011,Berninger2011,Ferlaino2009,Ferlaino2010,Machtey2012,Zenesini2012,Knoop2012}
have revealed Efimov trimers and thereby determined their three-body
parameters. In these experiments, rather broad Feshbach resonances
are used, implying that the range $b$ of the interactions between
atoms is typically the van der Waals length $r_{\mbox{\tiny vdW}}=\frac{1}{2}(mC_{6}/\hbar^{2})^{1/4}$
associated with the $-C_{6}/r^{6}$ tail of the open-channel potential~\citep{Chin2010}.
The measured value of the three-body parameter expressed in units
of $r_{\tiny\mbox{vdW}}$ turned out to stay fairly constant for different
atomic species~\citep{Ferlaino2010}, nuclear spin states~\citep{Gross2010},
or even different resonances of the same atomic species~\citep{Berninger2011}.
This indicates that the three-body parameter is universally determined
by the van der Waals length, and relatively insensitive to other short-range
details specific to individual atomic species.

It was first suggested that this van der Waals universality was due
to the very deep well of the potentials for these species, which support
many two-body bound states. According to this conjecture, when the
three atoms enter the short-range region of the potential, they feel
such a deep potential well that for all these species it results in
the same effect on the phase of the wave function, and leads to the
same three-body parameter~$\Lambda$. However, Efimov features for
helium atoms, which interact through a shallow potential supporting
only one two-body bound state, were shown to also follow the van der
Waals universality, both theoretically \citep{Naidon2012a} and experimentally~\citep{Knoop2012}.

A first attempt~\citep{Chin2011} to explain this universality suggested
that it could be due to quantum reflection in the sum of pairwise
$-1/r^{6}$ potentials. Particles coming from large distances to separations
on the order of the van der Waals length would experience a sudden
drop in the resulting effective three-body potential, which would
reflect the particles before they start to probe short-range physics.
However, a numerical study~\citep{JiaWang2012} showed that the relevant
three-body potential in the van der Waals region does not exhibit
a sudden drop, but a sudden repulsive barrier. The numerical results
indicate that this three-body repulsion is universally located around
$R\approx2\,r_{\tiny\mbox{vdW}}$ for several model potentials and
arises whenever a pairwise interaction potential features a deep well
supporting many two-body bound states or a short-range hardcore repulsion,
which is the case for all atomic species. Reference~\citep{JiaWang2012}
attributes the appearance of the repulsive barrier to an increase
of kinetic energy due to the squeezing of the hyperangular wave function
into a smaller volume caused by the suppression of two-body probability
inside the two-body potential well. Reference~\citep{Sorensen2012},
on the other hand, attributes it to the hard-core repulsion of the
two-body potential. It is therefore necessary to clarify how precisely
the repulsive barrier emerges, what physical picture it corresponds
to, and why it is universal. The purpose of this work is to show that
the universality of the three-body parameter indeed originates from
the two-body correlation, through a deformation of the three-body
system in the van der Waals region.\\

This paper is organised as follows. In Sec. II, we review how the
three-body repulsion setting the three-body parameter arises in the
hyperspherical formalism. In Sect.~III, we interpret the three-body
repulsion as a consequence of three-body deformation induced by pair
correlation and show why it is universal. In Sect.~IV, we confirm
this scenario by using two simple models. In Sect.~V, we give the
conclusion of this work.\\

\section{The three-body repulsion}

The three-body repulsion is observed in the hyperspherical formalism,
where the three-body wave function $\Psi$ is expressed in terms of
the hyperradius $R=\sqrt{\frac{2}{3}(r_{12}^{2}+r_{23}^{2}+r_{31}^{2})}$,
which corresponds to the global size of the three-body system~\footnote{There are several conventions for the definition of the hyperradius.
We choose the definition of Ref.~\citep{Efimov1971}, which differs
by a factor $\sqrt{\sqrt{3}/2}\approx0.93$ from that of Ref.~\citep{JiaWang2012}
for the case of three identical bosons.}, and the hyperangles, which describe the shape of the three-body
system and are collectively denoted by $\Omega$. We present here
a simple approximation that captures the bare essentials of both the
Efimov attraction and the universal three-body repulsion. 

The three-body wave function can be expanded over a basis of hyperangular
wave functions $\tilde{\Phi}_{n}(\Omega;R)$ which are normalised
to unity:
\begin{equation}
\Psi(R,\Omega)=\frac{1}{R^{5/2}}\sum_{n}f_{n}(R)\tilde{\Phi}_{n}(\Omega;R).\label{eq:TotalWaveFunction}
\end{equation}
The hyperangular wave function $\tilde{\Phi}_{n}$ itself can be expanded
into three Faddeev components $\phi_{n}^{(i)}$, 
\begin{equation}
\tilde{\Phi}_{n}(\Omega;\,R)=\sum_{i=1,2,3}\frac{\phi_{n}^{(i)}(\Omega;R)}{\sin2\alpha_{i}}.\label{eq:FaddeevDecomposition}
\end{equation}
Here,  $\alpha_{i}=\arctan\frac{\sqrt{3}r_{jk}}{2r_{i,jk}}=\arcsin\frac{r_{jk}}{R}$
is the Delves hyperangle in the $i$th Jacobi coordinate system $(\vec{r}_{jk},\vec{r}_{i,jk})$,
where $(i,j,k)$ denotes the cyclic permutations of (1,2,3). For identical
bosons, the functional forms of all Faddeev components are the same,
$\phi_{n}^{(i)}=\phi_{n}$. The advantage of this Faddeev decomposition
is that it treats the three particles on equal footing. In the low-energy
Faddeev approximation \citep{Fedorov1993}, $\phi_{n}$ is assumed
to depend only upon the Delves hyperangle $\alpha$. In this approximation,
one ignores the dependence of $\phi_{n}$ on the directions of $\vec{r}_{jk}$
and $\vec{r}_{i,jk}$, \emph{i.e.} higher angular momentum partial
waves. This excludes the possibility of accidental resonances with
higher partial waves \citep{DIncao2009,JiaWang2012}, and gives less
accurate results. Nevertheless, this approximation is good enough
for our purpose, as we shall see below.

The Faddeev component $\phi_{n}$ is chosen to be the eigensolution
with the eigenvalue $\lambda_{n}$ of the Faddeev equation
\begin{multline}
\left(-\frac{\partial^{2}}{\partial\alpha^{2}}-\lambda_{n}\right)\phi_{n}(\alpha)=\\
-R^{2}\frac{m}{\hbar^{2}}V(R\sin\alpha)\left(\phi_{n}(\alpha)+\frac{4}{\sqrt{3}}\int_{\alpha_{\tiny\mbox{min}}}^{\alpha_{\tiny\mbox{max}}}\phi_{n}(\alpha^{\prime})d\alpha^{\prime}\right),\label{eq:HyperangularEquation}
\end{multline}
where $m$ is the particle mass, $V$ is the pairwise interaction
potential, $\alpha_{\tiny\mbox{min}}=\vert\frac{\pi}{3}-\alpha\vert$,
and $\alpha_{\tiny\mbox{max}}=\frac{\pi}{2}-\vert\frac{\pi}{6}-\alpha\vert$.
We assume that three-body interactions are negligible~\footnote{This is justified to the extent that the range of the physical three-body
force does not exceed the range of the three-body repulsion.}.

Each solution $\phi_{n}$ defines a channel $n$, and one finds in
general that the hyperradial functions $f_{n}(R)$ are solutions of
the coupled equations,
\begin{multline}
\left(-\frac{\partial^{2}}{\partial R^{2}}+\frac{\lambda_{p}(R)}{R^{2}}-\frac{1}{4R^{2}}-\frac{m}{\hbar^{2}}E\right)f_{p}(R)\\
+\sum_{n}\left(Q_{pn}f_{n}(R)+2P_{pn}\frac{\partial f_{n}(R)}{\partial R}\right)=0,\label{eq:HyperradialEquation}
\end{multline}

with the nonadiabatic couplings
\begin{equation}
Q_{pn}=-\int d\Omega\tilde{\Phi}_{p}^{*}\frac{\partial^{2}\tilde{\Phi}_{n}}{\partial R^{2}};\quad\quad P_{pn}=-\int d\Omega\tilde{\Phi}_{p}^{*}\frac{\partial\tilde{\Phi}_{n}}{\partial R}.\label{eq:DefinitionOfPandQ}
\end{equation}

\begin{figure}
\includegraphics[viewport=0bp 0bp 270bp 211bp,clip,scale=0.9]{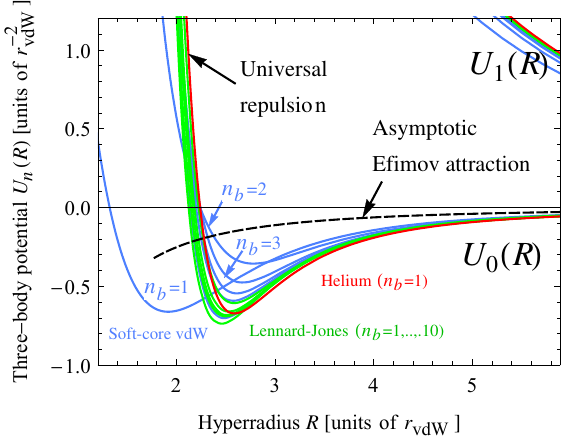}

\caption{\label{fig:ThreeBodyPotential}Three-body potentials $U_{0}(R)$ and
$U_{1}(R)$ for different pairwise interactions at unitarity: soft-core
van der Waals potential (blue) with $n_{b}=1$-10 two-body bound states,
Lennard-Jones potential (green) with $n_{b}=1$-10 two-body bound
states, and helium potential (red) rescaled to reach unitarity with
$n_{b}=1$ two-body bound state. Note that only the case of the soft-core
van der Waals potential with one bound state is significantly different
from the other cases. The dashed curve shows the asymptotic Efimov
attraction.}
\end{figure}

If we restrict our consideration to a particular channel $n$ (neglecting
couplings to other channels) and note that $P_{nn}=0$ due to the
normalisation of $\tilde{\Phi}_{n}$, we arrive at a simple Schrödinger
equation
\begin{equation}
\left(-\frac{\partial^{2}}{\partial R^{2}}-\frac{1}{4R^{2}}+U_{n}(R)-\frac{m}{\hbar^{2}}E\right)f_{n}(R)\approx0,\label{eq:SingleChannelEquation}
\end{equation}
with the three-body potential 
\begin{equation}
U_{n}(R)=\frac{\lambda_{n}(R)}{R^{2}}+Q_{nn}(R).\label{eq:ThreeBodyPotential}
\end{equation}
This potential is the sum of adiabatic (first term) and nonadiabatic
(second term) contributions. The Efimov attraction manifests itself
in this framework as the appearance in a particular channel $n=0$
of a negative eigenvalue $\lambda_{0}(R)\to-s_{0}^{2}$ at large hyperradii,
with $s_{0}\approx1.00624$. One can show that $Q_{00}(R)\to O(1/R^{3})$,
so that the potential $U_{0}(R)$ tends to the $1/R^{2}$ Efimov attraction
at large $R$. At shorter distance, the potential becomes repulsive.
This is illustrated in Fig.~\ref{fig:ThreeBodyPotential}, where
the three-body potentials $U_{n}(R)$ obtained by solving Eq.~(\ref{eq:HyperangularEquation})
are represented for several two-body potentials with a van der Waals
tail. Namely, we used the following soft-core van der Waals and Lennard-Jones
potentials,
\begin{eqnarray}
V_{\mbox{\tiny soft}}(r) & = & -C_{6}\frac{1}{r^{6}+\sigma^{6}},\label{eq:SoftVanderWaals}\\
V_{\tiny\mbox{LJ}}(r) & = & C_{6}\left(\frac{\sigma^{6}}{r^{12}}-\frac{1}{r^{6}}\right),\label{eq:LennardJones}
\end{eqnarray}
where the length $\sigma$ is adjusted to produce a shape resonance
(divergence of the scattering length, leading to the unitarity limit
$a\to\infty$). There are several possible choices of $\sigma$, corresponding
to different depths of the potential well, or equivalently different
numbers of $s$-wave two-body bound states $n_{b}$ (including the
one at the breakup threshold). We also use a realistic helium potential~\citep{Aziz1991}
rescaled to reach unitarity~\citep{Naidon2012a}, which is qualitatively
similar to a Lennard-Jones potential at unitarity with one two-body
bound state.

Figure~\ref{fig:ThreeBodyPotential} shows that for all these two-body
potentials, the three-body potential $U_{0}(R)$ in the Efimov channel
$(n=0)$ exhibits both the Efimov attraction at large distance and
a repulsive barrier at short distance. Consistent with Ref.~\citep{JiaWang2012},
for all these pairwise potentials with the exception of the soft-core
van der Waals potential with one bound state, the repulsive barrier
is universally located around $R\approx2r_{\mbox{\tiny vdW}}$.

\begin{figure}
\includegraphics[width=8.6cm]{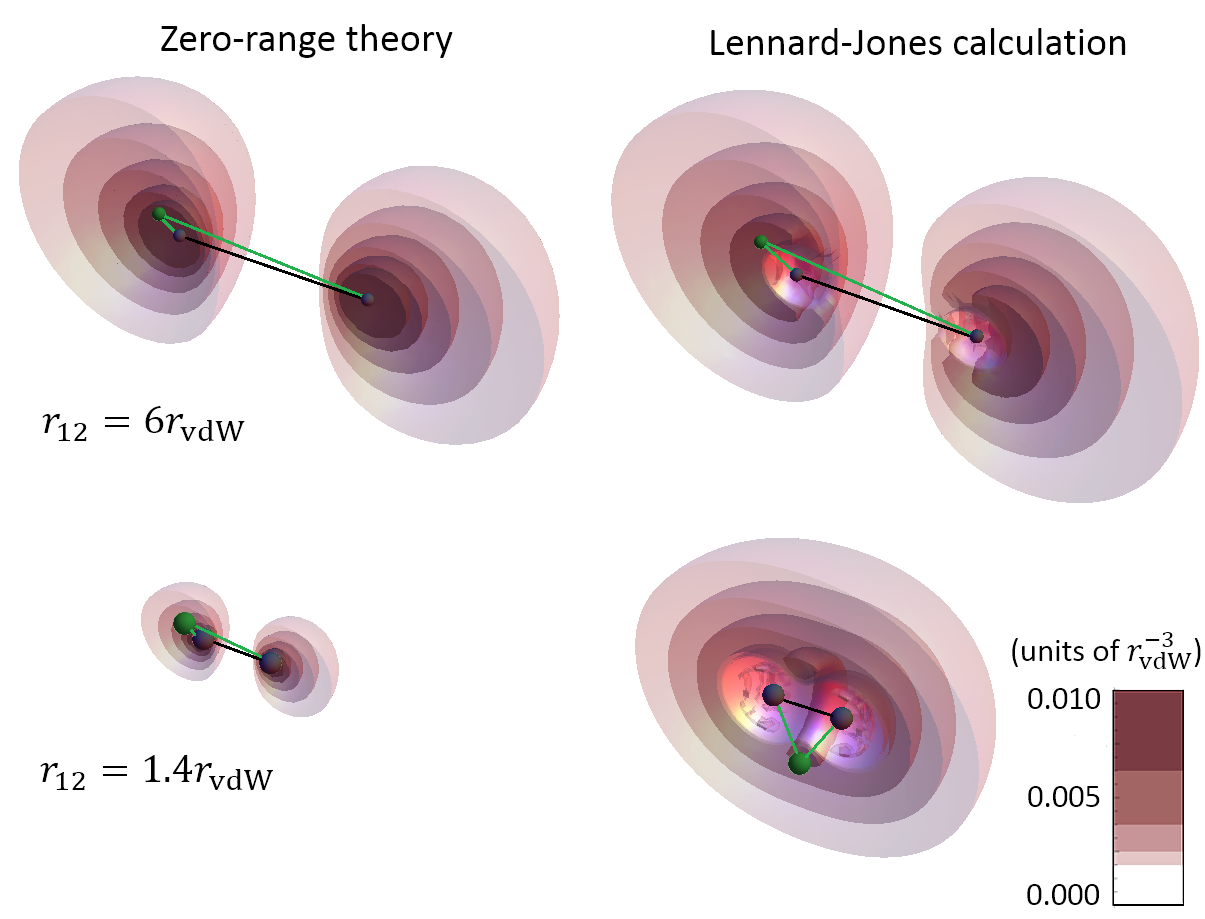}

\caption{\label{fig:ProbabilityDensity}Three-dimensional contour plots of
the probability distribution in Eq.~(\ref{eq:ProbabilityDensity})
of finding a particle for a given separation of the two other particles
(which are indicated by a pair of small gray balls connected by a
black line). For clarity, we only show the probability density behind
a plane containing the two particles, and shade the contours with
an opacity increasing with probability density: the darker, the higher
the probability of finding the third particle. The top figures correspond
to a separation of $6.0\,r_{\tiny\mbox{vdW}}$, while the bottom ones
corrrespond to a separation of $1.4r_{\tiny\mbox{vdW}}$. To appreciate
the change in configuration between the figures, a typical location
of the third particle is indicated by a small green ball connected
to the other two particles by green lines. The left figures were computed
from the zero-range Efimov theory at unitarity; they show the invariance
of the Efimov configuration distribution with respect to the size
of the system. The right figures were computed for a Lennard-Jones
pairwise potential at unitarity supporting four two-body bound states.
At large separations, the probability distribution is consistent with
the Efimov configuration distribution, but around each of the two
particles there is a noticeable sphere of radius $\sim r_{\tiny\mbox{vdW}}$
in which the probability is significantly suppressed. This suppression
leads to an abrupt change in configuration probability when the particles
come close.}
\end{figure}
\begin{figure}
\includegraphics[viewport=0bp 0bp 250bp 174bp,clip]{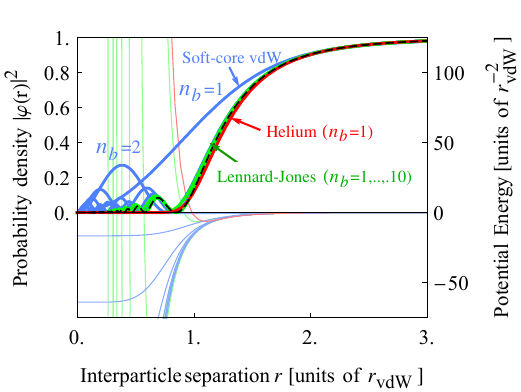}

\caption{\label{fig:TwoBody}Zero-energy two-body probability density distribution
$\vert\varphi\vert^{2}$ (normalised asymptotically to unity) as a
function of interparticle distance for different two-body potentials:
soft-core van der Waals potential (blue) with $n_{b}=1$-8 bound states,
Lennard-Jones potential (green) with $n_{b}=1$-8 bound states, helium
potential (red) rescaled to reach unitarity with one bound state.
The corresponding potentials are shown in faded colours. The probability
density corresponding to the universal van der Waals correlation given
in Eq.~(\ref{eq:universalCorrelation}) is shown by the dashed black
curve.}
\end{figure}

\section{Interpretation of the three-body repulsion}

\subsection{Repulsion due to deformation}

One might think that the repulsive barrier is a consequence of the
hard-core repulsion in the two-body potential, as suggested in Ref.~\citep{Sorensen2012}.
However, this is not the case since it occurs also for the soft van
der Waals potential which has no repulsive core and is purely attractive.
Another counter-intuitive observation is that the depth of the three-body
potential $U_{0}$ remains relatively stable as the two-body potential
is made deeper and deeper.  Our calculation shows that the adiabatic
contribution $\lambda_{0}(R)/R^{2}$ in Eq.~(\ref{eq:ThreeBodyPotential})
gets indeed deeper, but is compensated by the purely repulsive nonadiabatic
term $Q_{00}$. The fact that the nonadiabatic kinetic energy is indeed
repulsive at large distance can be understood by rewriting $Q_{00}$
as 
\begin{equation}
Q_{00}(R)=\int d\Omega\left|\frac{\partial\tilde{\Phi}_{0}(\Omega;R)}{\partial R}\right|^{2}\ge0,\label{eq:NonadiabaticKineticEnergy}
\end{equation}
using the normalisation of $\tilde{\Phi}_{0}$ and the fact that it
can be chosen to be real. This shows that $Q_{00}$ is positive and
since it has to vanish at large distance, it must be repulsive (if
one excludes unlikely oscillations at infinitely large distance).

Equation (\ref{eq:NonadiabaticKineticEnergy}) shows that the nonadiabatic
kinetic energy $Q_{00}$ arises from a change in the hyperangular
wave function $\tilde{\Phi}_{0}$ with respect to the hyperradius,
i.e. from a change in the probability distribution of the shape of
the three-body system as a function of its size. To visualise this
change, we use $\tilde{\Phi}_{0}$ to plot in Fig.~\ref{fig:ProbabilityDensity}
the probability density of finding a particle 3 for a given separation
$r_{12}$ of the two other particles 1 and 2, 
\begin{equation}
P(\vec{r}_{12,3})=(\sin2\alpha_{3})^{2}\left|\tilde{\Phi}_{0}(\Omega;R)\right|^{2}.\label{eq:ProbabilityDensity}
\end{equation}
When particle 3 is far from particles 1 and 2, the hyperangular wave
function is given by the zero-range limit (corresponding to the Efimov
theory), which at unitarity admits the following analytical solution~\citep{Efimov1971},
\begin{eqnarray}
\tilde{\Phi}_{0}^{\tiny\mbox{(ZR)}}(\Omega) & = & \sum_{i=1}^{3}\frac{\phi_{0}^{\tiny\mbox{(ZR)}}(\alpha_{i})}{\sin2\alpha_{i}},\nonumber \\
\mbox{with}\,\quad\phi_{0}^{\tiny\mbox{(ZR)}}(\alpha) & = & \sinh(s_{0}(\frac{\pi}{2}-\alpha)),\label{eq:ZeroRangeLimit}
\end{eqnarray}
which is independent of the hyperradius. The probability density therefore
remains the same up to a scale transformation. In other words, the
probability distribution of the shape of the three-particle system
remains the same, namely the third particle is typically located closer
to one of the other two. This invariance of the hyperangular wave
function with respect to the hyperradius results in $Q_{00}=0$.

When particle 3 comes close to particle 1 or 2, however, this zero-range
picture becomes invalid because the finite-range effects of the interaction
are is no longer negligible. In Fig.~\ref{fig:ProbabilityDensity},
one can clearly see two regions of suppressed probability near particles
1 and 2. This exclusion is an expected consequence of the known two-body
physics. It is expected indeed that for short-range interactions the
three-body density distribution becomes proportional to the relative
two-body density distribution whenever two particles come sufficiently
close, as recently illustrated in nuclear physics \citep{Alvioli2012}.
The relative radial probability density distribution $\vert\varphi\vert^{2}$
for two particles at zero scattering energy is represented in Fig.~\ref{fig:TwoBody}
for different two-body potentials at unitarity. One can see that the
probability is indeed suppressed below some radius on the order of
$r_{\mbox{\tiny vdW}}$ due to either the presence of a repulsive
wall or, on the contrary, the acceleration in the well of the potential.

As particles 1 and 2 come close, this two-body exclusion confines
the probability distribution for particle 3 to a region forming a
ring in between the two particles, corresponding to an equilateral
shape of the three-particle system. We find that this change of shape
happens very suddenly, making it difficult for the system to follow
the Efimov channel adiabatically (see the the supplementary materials
for the animations corresponding to the left side (zero-range theory)
and right side (Lennard-Jones calculations) of Fig. 2). This abrupt
variation results in a significant gain of nonadiabatic kinetic energy
$Q_{00}$ in Eq.~(\ref{eq:NonadiabaticKineticEnergy}), thereby creating
the three-body repulsion. In light of this discussion, we expect
the three-body repulsion to be essentially determined by the form
of the pair correlation.

\subsection{van der Waals universality}

The pair correlation for two particles interacting with van der Waals
interactions is known to have a universal asymptotic form~\citep{Gao1998,Flambaum1999}.
In particular, the zero-energy radial two-body wave function $\varphi$
for a given scattering length $a$ has the following analytical form
in the van der Waals tail region:

\begin{equation}
\varphi(r)=\Gamma(5/4)\sqrt{x}J_{\frac{1}{4}}\!\!\left(2x^{-2}\right)-\frac{r_{\tiny\mbox{vdW}}}{a}\Gamma(3/4)\sqrt{x}J_{-\frac{1}{4}}\!\!\left(2x^{-2}\right),\label{eq:universalCorrelation}
\end{equation}
where $\Gamma$ and $J_{\alpha}$ denote the gamma and Bessel functions,
and $x=\frac{r}{r_{\tiny\mbox{vdW}}}$. At large distance, $\varphi(r)$
asymptotes to the free wave form $1-\frac{r}{a}$. For $a\to\infty$,
$\varphi(r)$ asymptotes to unity and thus can be regarded as a correlation
function describing the deviations from the free wave. The corresponding
two-body probability density $\vert\varphi\vert^{2}$ is represented
in Fig.~\ref{fig:TwoBody} by the black dashed curve. One can see
that the probability densities obtained for all the considered potentials
nearly coincide with this analytical form for $r\gtrsim r_{\tiny\mbox{vdW}}$.
For potentials which strongly suppress the probability for $r\lesssim r_{\tiny\mbox{vdW}}$,
the whole pair correlation is thus very similar to the universal correlation.
This similarity holds even for shallow potentials with a short van
der Waals tail accompanied by a hard-core repulsion, such as that
of helium. The fact that the short-distance oscillations of the universal
correlation are not reproduced does not make any major difference,
because the probability density is very small in this region. Since
the two-body correlation is nearly the same for these potentials,
the same nonadiabatic deformation occurs, leading to the same three-body
repulsion and three-body parameter. Conversely, the soft-core van
der Waals potential with one two-body bound state leads to a pair
correlation that deviates from the universal correlation more significantly,
with a less pronounced suppression of probability, as seen in Fig.~\ref{fig:TwoBody}.
According to our interpretation, this should create a softer three-body
repulsion at a shorter hyperradius. This is indeed the case, as can
be checked in Fig.~\ref{fig:ThreeBodyPotential}.

In the present interpretation, the universality of the three-body
parameter is thus a direct consequence of the van der Waals two-body
correlation.

\section{Check with simple models}

To verify our interpretation, we construct two simple models. The
first one verifies that the pair correlation does indeed create a
three-body repulsive barrier at $R\approx2r_{vdW}$ in the Efimov
channel through the nonadiabatic kinetic energy. The second one is
a more complete model verifying quantitatively that the pair correlation
fixes the three-body parameter to a value consistent with full numerical
calculations and experiments. 

\subsection{Pair correlation model}

\begin{figure}[tph]
\includegraphics[viewport=0bp 0bp 190bp 159bp,clip,scale=0.7]{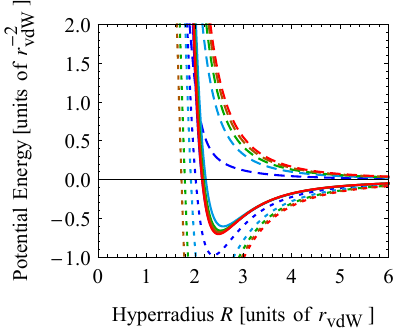}\includegraphics[viewport=45bp 0bp 190bp 159bp,clip,scale=0.7]{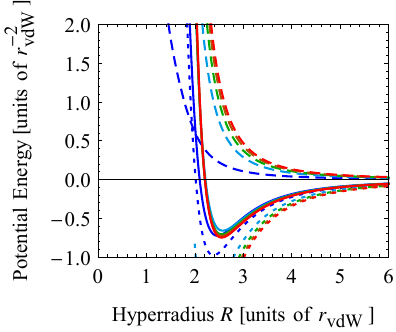}

\caption{\label{fig:Comparison}Comparison between the Faddeev three-body calculations
(left) and the simple two-body correlation model described in the
main text (right). The dashed curves show the nonadiabatic kinetic
energy $Q_{00}$ for Lennard-Jones potentials of different depths,
corresponding to the unitarity limit with different numbers of two-body
bound states ranging from 1 to 5. The solid curves show the full three-body
potential $U_{0}(R)$ obtained by adding to $Q_{00}$ the adiabatic
contribution $\lambda_{0}/R^{2}$ obtained from Faddeev calculations,
which is shown by the dotted curves. }
\end{figure}
To simply account for the two-body suppression, we consider a trial
hyperangular wave function of the Bijl-Jastrow form \citep{Bijl1940,Jastrow1955},
which is the uncorrelated hyperangular function $\tilde{\Phi}_{0}^{\tiny\mbox{(ZR)}}$
in the zero-range (Efimov) limit given by Eq.~(\ref{eq:ZeroRangeLimit}),
multiplied by a product of the universal two-body correlation $\varphi$
given by Eq.~(\ref{eq:universalCorrelation}), which causes the suppression
of probability in the two-body sector:
\begin{equation}
\tilde{\Phi}_{0}^{\tiny\mbox{(model)}}=\tilde{\Phi}_{0}^{\tiny\mbox{(ZR)}}\times\prod_{i<j}\varphi(r_{ij}).\label{eq:Ansatz}
\end{equation}

This simple ansatz leads to a probability density that is very similar
to the one calculated from the Faddeev equation (\ref{eq:HyperangularEquation}).
In particular, we have confirmed that it also leads to a sudden buildup
of probability in the ring-shaped region when two particles are close.
One can also calculate the nonadiabatic kinetic energy $Q_{00}$ from
Eq.~(\ref{eq:NonadiabaticKineticEnergy}). As expected, we find a
sudden increase of $Q_{00}$ at the hyperradius $R\approx2r_{\mbox{\tiny vdW}}$.
This model thus confirms our claim that the nonadiabatic change in
configuration originates from an interplay between the suppression
of two-body probability and the Efimov configuration.

Adding the adiabatic term $\frac{\lambda_{0}}{R^{2}}$ to $Q_{00}$,
we obtain the full potential $U_{0}(R)$. As shown in Fig.~\ref{fig:Comparison},
it reproduces very well the universal potential found using the solution
of the Faddeev equation (\ref{eq:HyperangularEquation}). Note that
this agreement is remarkable; although the adiabatic and nonadiabatic
terms taken separately vary signicantly for different numbers of two-body
bound states, their variations almost cancel out to give the universal
potential. 

\subsection{Separable model}

The hyperspherical formalism is useful to exhibit the three-body repulsion
mechanism, and the previous model satisfactorily reproduces the three-body
repulsion in the Efimov channel. However, this channel alone only
gives qualitative results for the actual trimer energies. To be more
quantitative, one would need to solve the many coupled equations in
Eq. (\ref{eq:HyperradialEquation}), as done in Ref.~\citep{JiaWang2012},
but that would defeat our purpose of using a simple model to reproduce
the physics. Hence we turn to another approach to get more quantitative
results, while keeping the central idea of the universal pair correlation
in Eq.~(\ref{eq:universalCorrelation}) being the essential ingredient
behind the universal three-body parameter.

One of the simplest pseudo-potentials that can reproduce the universal
pair correlation is the separable potential~\citep{Yamaguchi1954,EST1973},
\begin{equation}
\hat{V}=\frac{\hbar^{2}}{m}\xi\vert\chi\rangle\langle\chi\vert,\label{eq:SeparablePotential}
\end{equation}
where the function $\chi$ in momentum space is chosen to be
\begin{equation}
\chi(q)=1-q\int_{0}^{\infty}dr\left(1-\frac{r}{a}-\varphi(r)\right)\sin(qr),\label{eq:Chi}
\end{equation}
and the coefficent $\xi$ is set to
\begin{equation}
\xi=4\pi\left(\frac{1}{a}-\frac{2}{\pi}\int_{0}^{\infty}dq\vert\chi(q)\vert^{2}\right)^{-1}.\label{eq:Xi}
\end{equation}
This potential has the advantage of being easily tractable because
of its separability, and one can show (see Appendix B) that the solution
of the two-body problem at zero energy for this potential is given
exactly by $\varphi(r)$, which is chosen to be the universal pair
correlation given by Eq.~(\ref{eq:universalCorrelation}). Numerically,
we find that this potential is an excellent substitute for the real
van der Waals interaction in the two-body problem at low energy: it
reproduces the low-energy scattering state and the two-body bound
state over energies on the order of $\hbar^{2}/(mr_{\tiny\mbox{vdW}}^{2})$
and scattering lengths $\vert a\vert\gtrsim2r_{\tiny\mbox{vdW}}$. 

For the three-body problem, substituting the real potential by the
separable potential in the three-body Schrödinger equation leads to
a one-dimensional integral equation \citep{Lee2007,Naidon2011} that
is similar to the Skorniakov--Ter-Martirosian equation obtained for
a contact potential~\citep{Skorniakov1957} and that can easily be
solved numerically (see the Appendix B for the derivation). We emphasise
that the only information contained in this model is the zero-energy
pair correlation.

\begin{figure}
\includegraphics[scale=0.9]{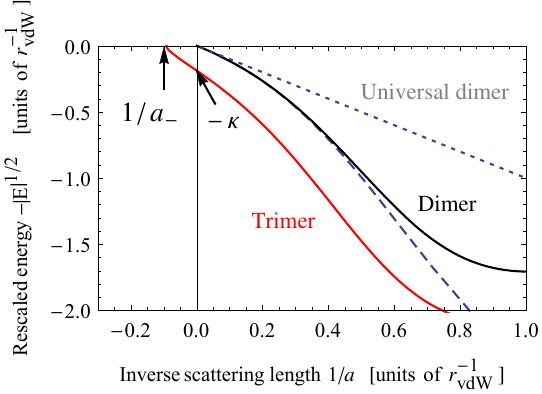}

\caption{\label{fig:SeparableSpectrum}Lowest trimer and dimer energy for the
separable potential given by Eq.~(\ref{eq:SeparablePotential}),
as a function of its inverse scattering length $1/a$. For comparison,
the dotted and dashed curves represent the universal dimer energy
($E=-\frac{\hbar^{2}}{ma^{2}}$) and the exact van der Waals dimer
energy, respectively. Both the abcissa and ordinate are shown in units
of the inverse van der Waals length $r_{\mbox{\tiny vdW}}^{-1}$.}
\end{figure}

From the numerical solution, we obtain the ground-state trimer spectrum
shown in Fig.~\ref{fig:SeparableSpectrum}. In particular, we extract
the binding wave number $\kappa$ at unitarity and the scattering
length $a_{-}$ at which this trimer disappears in the three-body
threshold, and find
\[
\kappa\,r_{\tiny\mbox{vdW}}=0.187(1)\qquad\mbox{and}\qquad\frac{a_{-}}{r_{\tiny\mbox{vdW}}}=-10.86(1).
\]

\begin{figure}
\includegraphics[width=8cm]{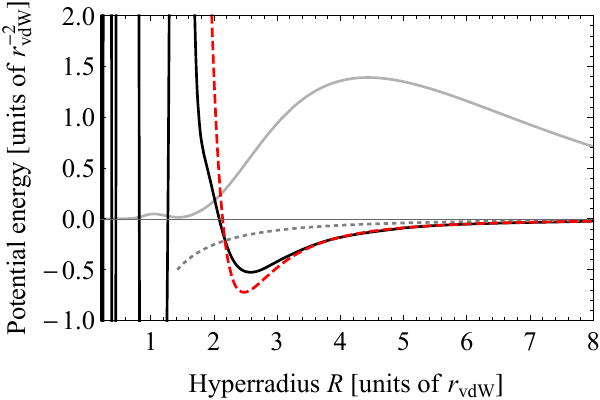}

\caption{\label{fig:Potential}Three-body probability (solid gray curve, in
arbitrary units) as a function of hyperradius, and corresponding effective
three-body potential (solid black curve), obtained from the separable
model Eq.~(\ref{eq:SeparablePotential}). For comparison, the Efimov
potential (dotted curve) and the low-energy Faddeev three-body potential
obtained from Eq.~(\ref{eq:ThreeBodyPotential}) for a Lennard-Jones
potential with six s-wave bound states (red dashed curve) are also
represented. Note that for $R<1.5r_{\mbox{\tiny vdW}}$, the effective
potential becomes strongly oscillatory (as can be seen from the nearly
vertical lines) in order to reproduce the wave function. Since the
physics in that region is correctly described by coupled potentials,
the effective potential is not a meaningful construct in that region
and the short-range oscillations have no particular significance.}

\end{figure}
Consistent with our interpretation, similar results are obtained for
pair correlations $\varphi(r)$ with a similar tail and low probability
at short distance. For instance, the pair correlation for a Lennard-Jones
potential with one two-body bound state leads to $\kappa\,r_{\tiny\mbox{vdW}}=0.205(1)$
and $\frac{a_{-}}{r_{\tiny\mbox{vdW}}}=-10.23(1)$. These values agree
within a few percent with the results of \citep{JiaWang2012}, and
are in fair agreement with the experimental result of $\frac{a_{-}}{r_{\tiny\mbox{vdW}}}=-9.1(5)$~\citep{Ferlaino2011}. 

Finally, one can check that this agreement is not coincidental, as
the model also reproduces the deformation and repulsion effects. This
is demonstrated by the three-body probability  plotted as a function
of $R$ in Fig.~\ref{fig:Potential}. By calculating the second-order
derivative of the corresponding wave function, one can derive an effective
hyperradial potential (see Appendix~C). This potential reproduces
very closely the three-body potential calculated from the low-energy
Faddeev equation. Only the repulsive barrier is slightly shifted to
a smaller hyperradius, resulting in a more quantitatively accurate
three-body parameter.

\section{Conclusion}

We have shown that the universality of the three-body parameter revealed
in recent experiments with neutral atoms and numerical calculations
originates from two-body correlation.

The mechanism explaining this origin is the following: two-body correlation
suppresses the probability for two atoms to be at separations smaller
than the van der Waals length, which imposes a deformation of the
three-atom system when the three atoms come withing the distance of
the van der Waals length. The kinetic-energy cost associated with
this deformation creates a repulsion preventing the three atoms from
coming closer, and sets the three-body parameter.

This mechanism is consistent with the findings of J.~Wang \emph{et
al.}~\citep{JiaWang2012}. Unlike other proposed mechanisms~\citep{Sorensen2012,Chin2011},
this does not necessitate a hard-core repulsion in the two-body potential,
nor is it simply the expression of quantum reflection along a single
coordinate since it involves the three-body deformational degrees
of freedom.

Because the two-body correlation is universally determined by the
van der Waals length for atomic systems, this makes the three-body
parameter universal in these systems. More generally, we expect to
find such universality in any class of systems where the two-body
suppression has a universal form. This work also suggests that for
other systems, in general, pair correlations and their associated
length scale, the effective range, should play an essential role.
These points are addressed in more detail in a separate paper~\citep{Naidon2014}.\\

We thank C. Greene for suggesting the plot in Fig.~\ref{fig:Potential}
and Ch. Elster for making us aware of Ref.~\citep{EST1973}. P.
N. acknowledges support from RIKEN through the Incentive Research
Project funding. S. E. acknowledges support from JSPS. M. U. and P.
N. acknowledge the hospitality of the Aspen Center for Physics where
part of this work was done. M. U. acknowledges the financial support
from Grants-in-Aid (KAKENHI Grant Nos. 26287088 and 22103005) and
the Photon Frontier Network Program from MEXT of Japan.

\section*{}

\section*{Appendix A}

In this Appendix, we show that the radial wave function $\varphi$
in Eq.~(\ref{eq:Chi}) is the solution of the two-body problem at
zero energy for the separable potential given by Eq.~(\ref{eq:SeparablePotential}).
A more general representation of two-body interactions in terms of
separable potentials can be found in Ref.~\citep{EST1973}.

The two-body Schrödinger equation at zero energy in momentum space
reads:
\begin{equation}
\frac{\hbar^{2}p^{2}}{m}\tilde{\psi}(\vec{p})+\int\frac{d^{3}\vec{q}}{(2\pi)^{3}}\tilde{V}(\vec{p},\vec{q})\tilde{\psi}(\vec{q})=0,\label{eq:AppendixA1}
\end{equation}
where $\tilde{V}$ is the Fourier transform of the pairwise potential~$V$,
and $\tilde{\psi}$ is the Fourier transform of the two-body wave
function. Replacing $\tilde{V}$ by the separable potential in Eq.~(\ref{eq:SeparablePotential}),
one obtains: 
\begin{equation}
p^{2}\tilde{\psi}(\vec{p})-f\,\chi(p)=0,\label{eq:AppendixA2}
\end{equation}
with
\begin{equation}
f=-\int\frac{q^{2}dq}{2\pi^{2}}\tilde{\psi}(\vec{q})\xi\chi^{*}(q).\label{eq:AppendixA3}
\end{equation}
Inserting Eq.~(\ref{eq:AppendixA2}) into Eq.~(\ref{eq:AppendixA3}),
one obtains the explicit expression for $f$:
\begin{equation}
f=-\left(\frac{1}{\xi}+\int\frac{dq}{2\pi^{2}}\vert\chi(q)\vert^{2}\right)^{-1},\label{eq:AppendixA4}
\end{equation}
and using the chosen form of $\xi$ given by Eq.~(\ref{eq:Xi}),
one obtains $f=-4\pi a$. Inserting this value into Eq.~(\ref{eq:AppendixA2}),
and inverting the resulting equation with the proper boundary conditions,
one finds:
\begin{equation}
\tilde{\psi}(\vec{p})=(2\pi)^{3}\delta^{3}(\vec{p})-4\pi a\frac{\chi(p)}{p^{2}}.\label{eq:AppendixA5}
\end{equation}
This translates in space coordinates as
\begin{equation}
\psi(\vec{r})=1-4\pi a\int\frac{d^{3}p}{(2\pi)^{3}}\frac{\chi(p)}{p^{2}}e^{i\vec{p}\cdot\vec{r}},\label{eq:AppendixA6}
\end{equation}
which after angular integration yields:
\begin{equation}
\psi(\vec{r})=1-a\frac{2}{\pi}\int_{0}^{\infty}\chi(p)\frac{\sin pr}{pr}dp.\label{eq:AppendixA7}
\end{equation}
Inserting the chosen form of $\chi$ given by Eq.~(\ref{eq:Chi}),
and using the closure relation $\int_{0}^{\infty}dp\sin(pr^{\prime})\sin(pr)=\frac{\pi}{2}\delta(r-r^{\prime})$,
one obtains
\begin{equation}
\psi(\vec{r})=-\frac{a}{r}\varphi(r),\label{eq:AppendixA8}
\end{equation}
which shows that the form of the zero-energy radial wave function
$r\psi(\vec{r})$ is indeed given by $\varphi(r)$.

\section*{Appendix B}

In this Appendix, we derive the equation we use to solve the three-body
problem with the separable potential given by Eq.~(\ref{eq:SeparablePotential}).

The three-body Schrödinger equation in momentum space reads:
\begin{multline}
\left(\frac{3}{4}\frac{\hbar^{2}}{m}P^{2}+\frac{\hbar^{2}}{m}p^{2}-E\right)\tilde{\Psi}(\vec{P},\vec{p})\\
+\sum_{i=1,2,3}\int\!\!\frac{d^{3}\vec{q}_{i}}{(2\pi)^{3}}\tilde{V}(\vec{p}_{i},\vec{q}_{i})\tilde{\Psi}(\vec{P}_{i},\vec{q}_{i})=0,\label{eq:AppendixB1}
\end{multline}
where $\tilde{V}$ is the Fourier transform of the pairwise potential~$V$,
and $\tilde{\Psi}$ is the Fourier transform of the three-body wave
function $\Psi$ in Eq.~(\ref{eq:TotalWaveFunction}) expressed in
a particular Jacobi coordinate set $(\vec{P},\vec{p})$ chosen among
the three possible sets $(\vec{P}_{i},\vec{p}_{i})$ with $i=1,2,3$.

Substituting $\tilde{V}$ by the separable potential in Eq.~(\ref{eq:SeparablePotential}),
one obtains: 
\begin{equation}
\left(\frac{3}{4}P^{2}+p^{2}-\frac{m}{\hbar^{2}}E\right)\tilde{\Psi}(\vec{P},\vec{p})+\sum_{i=1,2,3}F(\vec{P}_{i})\chi(p_{i})=0,\label{eq:AppendixB2}
\end{equation}
where 
\begin{equation}
F(\vec{P})=\xi\int\!\!\frac{d^{3}\vec{p}}{(2\pi)^{3}}\chi^{*}(p)\tilde{\Psi}(\vec{P},\vec{p}).\label{eq:AppendixB3}
\end{equation}

For $E<0$, Eq.~(\ref{eq:AppendixB2}) can be inverted as
\begin{equation}
\tilde{\Psi}(\vec{P},\vec{p})=-\sum_{i=1,2,3}\frac{F(\vec{P}_{i})\chi(p_{i})}{\frac{3}{4}P^{2}+p^{2}-\frac{m}{\hbar^{2}}E},\label{eq:AppendixB4}
\end{equation}
Inserting Eq.~(\ref{eq:AppendixB4}) into Eq.~(\ref{eq:AppendixB3})
gives:
\begin{equation}
\frac{1}{\xi}F(\vec{P})=-\sum_{i=1,2,3}\int\!\!\frac{d^{3}\vec{p}}{(2\pi)^{3}}\chi^{*}(p)\frac{F(\vec{P}_{i})\chi(p_{i})}{\frac{3}{4}P^{2}+p^{2}-\frac{m}{\hbar^{2}}E}.\label{eq:AppendixB5}
\end{equation}
Making the choice $(\vec{P},\vec{p})=(\vec{P}_{3},\vec{p}_{3})$,
one can factorise one of the terms in the sum with the left-hand side
of Eq.~(\ref{eq:AppendixB5}) as follows:
\begin{multline}
\left(\frac{1}{\xi}+\int\!\!\frac{d^{3}\vec{p}}{(2\pi)^{3}}\frac{\vert\chi(p)\vert^{2}}{\frac{3}{4}P^{2}+p^{2}-\frac{m}{\hbar^{2}}E}\right)F(\vec{P})\\
+\sum_{i=1,2}\int\!\!\frac{d^{3}\vec{p}}{(2\pi)^{3}}\chi^{*}(p)\frac{F(\vec{P}_{i})\chi(p_{i})}{\frac{3}{4}P^{2}+p^{2}-\frac{m}{\hbar^{2}}E}=0.\label{eq:AppendixB6}
\end{multline}
The two remaining terms are equal due to bosonic exchange symmetry,
and expressing one Jacobi coordinate set in terms of another, one
finally arrives at the integral equation for $F$: 
\begin{multline}
\left(\frac{1}{\xi}+\int\!\!\frac{d^{3}\vec{q}}{(2\pi)^{3}}\frac{\vert\chi(q)\vert^{2}}{q^{2}-(\frac{mE}{\hbar^{2}}-\frac{3}{4}P^{2})}\right)F(\vec{P})\\
+2\int\frac{d^{3}\vec{q}}{(2\pi)^{3}}\frac{\chi^{*}(\left|\vec{q}+\frac{\vec{P}}{2}\right|)\chi(\left|\frac{\vec{q}}{2}+\vec{P}\right|)}{P^{2}+q^{2}+\vec{q}\cdot\vec{P}-\frac{mE}{\hbar^{2}}}F(\vec{q})=0.\label{eq:AppendixB7}
\end{multline}
For spherically-symmetric solutions, it can be reduced to an equation
in which $F$ depends only on the one-dimensional variable $P=\vert\vec{P}\vert$:
\begin{equation}
D(P)F(P)+\int_{0}^{\infty}\frac{q^{2}dq}{2\pi^{2}}H(P,q)F(q)=0,\label{eq:AppendixB8}
\end{equation}
 with
\begin{equation}
D(P)=\frac{1}{\xi}+\int_{0}^{\infty}\frac{dq}{2\pi^{2}}\frac{q^{2}\vert\chi(q)\vert^{2}}{q^{2}-(\frac{mE}{\hbar^{2}}-\frac{3}{4}P^{2})},\label{eq:AppendixB9}
\end{equation}
\begin{equation}
H(P,q)=\int_{-1}^{1}\!du\frac{\chi^{*}(\sqrt{q^{2}\!+\!\frac{1}{4}P^{2}\!+\!qPu})\chi(\sqrt{P^{2}\!+\!\frac{1}{4}q^{2}\!+\!qPu})}{P^{2}+q^{2}+qPu-\frac{mE}{\hbar^{2}}}.\label{eq:AppendixB10}
\end{equation}

Solving for the eigenvalues of the linear operator in the left-hand
side of Eq.~(\ref{eq:AppendixB8}) and looking for the energies $E$
that make one of these eigenvalues equal to zero, consistent with
the right-hand side of Eq.~(\ref{eq:AppendixB8}), yields the energies
of three-body bound states. The corresponding eigenvectors $F$ give
the three-body wave functions $\Psi$ through Eq.~(\ref{eq:AppendixB4}).

\section*{Appendix C}

The explicit procedure to calculate the effective hyperradial potential
(black curve of Fig.~7) is the following.

First, the form of the Faddeev components in hyperspherical coordinates
is calculated from the solution $F(P)$ of Eq.~(33) in Appendix~B:
\begin{align}
\mathcal{F}(R,\alpha) & =\frac{-4}{\sqrt{3}R^{2}\sin2\alpha}\int\int\frac{PdP}{2\pi^{2}}\frac{pdp}{2\pi^{2}}\frac{F(P)\chi(p)}{\frac{3}{4}P^{2}+p^{2}-\frac{m}{\hbar^{2}}E}\nonumber \\
 & \qquad\times\sin(P\frac{\sqrt{3}}{2}R\cos\alpha)\sin(pR\sin\alpha).\label{eq:FaddeevComponent}
\end{align}

The total wave function is thus
\begin{equation}
\Psi(R,\alpha)=\mathcal{F}(R,\alpha_{3})+\mathcal{F}(R,\alpha_{2})+\mathcal{F}(R,\alpha_{1})\label{eq:TotalWavefunction}
\end{equation}
with 
\begin{align}
\alpha_{3} & =\alpha,\label{eq:alpha3}\\
\alpha_{2} & =\frac{1}{2}\left(\pi-\arccos\left(\frac{1}{2}\cos2\alpha-u\frac{\sqrt{3}}{2}\sin2\alpha\right)\right),\label{eq:alpha2}\\
\alpha_{1} & =\frac{1}{2}\left(\pi-\arccos\left(\frac{1}{2}\cos2\alpha+u\frac{\sqrt{3}}{2}\sin2\alpha\right)\right),\label{eq:alpha1}
\end{align}
where $u$ is the cosine of the angle between the two Jacobi vectors
$\vec{r}_{12}$ and $\vec{r}_{3,12}$. The hyperradial density (grey
curve of Fig.~7) is therefore given by:
\begin{equation}
\rho(\mathcal{R})\propto R^{5}\int_{-1}^{1}du\int_{0}^{\pi/2}\sin^{2}2\alpha\vert\Psi(R,\alpha)\vert^{2}.\label{eq:rho}
\end{equation}

Interpreting this density as an effective hyperradial wave function
$f(R)=\sqrt{\rho(R)}$ satisfying a Schrödinger equation similar to
Eq.~(\ref{eq:SingleChannelEquation}) for an uncoupled potential
$U_{\text{eff}}(R)$, one finds:
\begin{equation}
U_{\text{eff}}(R)=\frac{1}{4R^{2}}+\frac{f^{\prime\prime}(R)}{f(R)}+\frac{m}{\hbar^{2}}E.\label{eq:Ueff}
\end{equation}

\bibliographystyle{apsrev}
\bibliography{paper18e}

\end{document}